\documentclass[lettersize,journal]{IEEEtran}
\IEEEoverridecommandlockouts

\usepackage{fix-cm}
\usepackage{cite}
\usepackage{amsmath,amssymb,amsfonts}
\usepackage{algorithmic}
\usepackage{graphicx,color}
\usepackage{textcomp}
\usepackage{hyperref}
\usepackage{graphicx}
\usepackage{units}
\usepackage{booktabs}
\usepackage{multirow}

\usepackage[caption=false,font=footnotesize]{subfig}

\usepackage{enumitem}

\newcommand{\abs}[1]{\left|#1\right|}
\newcommand{\mhyphen}{\mathord{-}}
\newcommand{\norm}[1]{\left\|{#1}\right\|}

\usepackage[cmintegrals]{newtxmath}

\makeatletter
\def\BState{\State\hskip-\ALG@thistlm}
\makeatother

\newcommand{\nth}[1]{{#1}{\text{th}}}

\newcommand{\mbf}[1]{\mathbf{#1}}
\newcommand{\Strpow}{{\sf *}}
\newcommand{\Hpow}{{\sf H}}
\newcommand{\Tpow}{{\sf T}}
\newcommand{\Invpow}{{\sf -1}}

\makeatother

\makeatletter
\def\ps@IEEEtitlepagestyle{%
  \def\@oddfoot{\mycopyrightnotice}%
  \def\@oddhead{\hbox{}\@IEEEheaderstyle\leftmark\hfil\thepage}\relax
  \def\@evenhead{\@IEEEheaderstyle\thepage\hfil\leftmark\hbox{}}\relax
  \def\@evenfoot{}%
}
\makeatother

\def\mycopyrightnotice{%
  \begin{minipage}{\textwidth}
  \centering \scriptsize
    This work has been accepted by the IEEE Open Journal of the Communications Society for publication.  Copyright may be transferred without notice, after which this version may no longer be accessible.
  \end{minipage}
}

\begin{document}

\title{Diversity Analysis for Terahertz Communication Systems under Small-Scale Fading}
 
\author{
        Almutasem Bellah Enad, \IEEEmembership{Graduate Student Member, IEEE,}
        Jihad Fahs, \IEEEmembership{Member, IEEE,}
        Hakim Jemaa, \IEEEmembership{Graduate Student Member, IEEE}
        Hadi Sarieddeen, \IEEEmembership{Senior Member, IEEE,}
        and Tareq Y. Al-Naffouri, \IEEEmembership{Fellow Member, IEEE}%
        
\thanks{This work was supported in part by AUB’s University Research Board and the Vertically Integrated Projects Program, and in part by KAUST’s Office of Sponsored Research under Award ORFS-CRG12-2024-6478.

Almutasem Bellah Enad, Jihad Fahs, and Hadi Sarieddeen are with the American University of Beirut (AUB), Beirut 1107--2020, Lebanon (e-mail: aae118@mail.aub.edu; jihad.fahs@aub.edu.lb; hadi.sarieddeen@aub.edu.lb).
Hakim Jemaa and Tareq Y. Al-Naffouri are with the King Abdullah University of Science and Technology (KAUST), Thuwal 23955--6900, Saudi Arabia (e-mail: hakim.jemaa@kaust.edu.sa; tareq.alnaffouri@kaust.edu.sa).
}}
\maketitle

\begin{abstract}
The terahertz (THz) band is a key enabler for future wireless systems, promising ultra-high data rates and dense spatial reuse. However, the reliability of THz links remains a major challenge due to severe path loss and small-scale fading effects, particularly in dynamic indoor and outdoor environments. This paper presents a comprehensive diversity analysis framework for THz communication systems under small-scale fading conditions. We model fading statistically using the generalized $\alpha$-$\mu$ distribution for indoor scenarios and the mixture of gamma (MG) model for outdoor propagation. We complement previous works that analyzed diversity under the $\alpha$-$\mu$ channels~\cite{10003243,9250676}. In particular, we present new insights on diversity for the MG channel in addition to recovering the results of~\cite{9250676} using a different approach. Moreover, we derive asymptotic expressions for the bit error rate as a function of the inverse signal-to-noise ratio, recovering all of the $\alpha$-$\mu$ diversity results using a simpler approximation method. The analytical results are extensively validated through Monte Carlo simulations, demonstrating excellent agreement. Our findings show that diversity gains in THz systems are strongly influenced by the number of independent paths, the severity of fading, and frequency selectivity. The proposed framework provides system designers with clear guidelines for quantifying and optimizing diversity gains in emerging channel models, paving the way for more reliable high-frequency wireless links in next-generation networks.
\end{abstract}


\maketitle

\section{INTRODUCTION}
\IEEEPARstart{T}{he} terahertz (THz) band, which spans 100 \,GHz to 10 \,THz, is poised to enhance future wireless systems by enabling ultra-high data rates in the terabit-per-second range and supporting extremely low-latency communications~\cite{sarieddeen2019generation,sarieddeen2020overview}. These capabilities are essential for emerging applications such as real-time holographic communication, high-resolution sensing, and ultra-reliable machine-type communication~\cite{rajatheva2020white,Jornet2024Evolution}. Despite these promises, THz propagation faces unique physical challenges. Recent studies reveal that THz signals exhibit significant temporal and spatial sparsity~\cite{sheikh2022thz,9591285}, yet can still support non-line-of-sight (NLoS) communication under favorable conditions~\cite{sheikh2022thz}. However, in practice, even in indoor environments, the availability of NLoS paths is limited and becomes further constrained under high-gain directional antennas~\cite{9591285}, or under high beamforming gains in multiple-input multiple-output (MIMO) systems~\cite{Faisal9216613}, which are often necessary to overcome severe THz-band path loss. In this context, accurately characterizing small-scale fading becomes critical to ensuring link reliability and efficient resource allocation. Such accurate modeling becomes even more critical under the constraints imposed by low-complexity signal processing techniques for THz-band baseband implementations \cite{Sarieddeen2024Bridging,Jemaa10757607}.

Measurement-based analyzes have shown that the generalized $\alpha$-$\mu$ distribution offers a robust statistical model to describe small-scale fading behavior in indoor THz channels~\cite{11072420}, making it a powerful tool for performance prediction and system design. Unlike classical fading models such as the Nakagami-$m$ and Rician distributions, which often fall short in capturing the complex and highly correlated nature of THz propagation, the $\alpha$-$\mu$ model provides a flexible and accurate parameterization tailored for such environments~\cite{papasotiriou2021experimentally}. Its versatility has proven effective in building analytical frameworks for indoor THz communication, particularly when combined with misalignment fading effects modeled via the zero-boresight approach~\cite{8610080}. These enhancements are critical for supporting reliable beamforming and alignment strategies in dense network deployments. 

The MG distribution, originally introduced for symbol error analysis in conventional wireless systems~\cite{6059452}, has also been successfully applied to characterize diversity reception over generalized-$K$ fading channels~\cite{6839048}. Its flexibility in approximating a wide range of fading conditions makes it a suitable candidate for modeling outdoor THz channels in next-generation networks. Recent works have further explored its application in THz environments. For example,~\cite{Jemaa2024Performance} presented a comprehensive performance analysis framework for outdoor THz SISO channels under MG fading with beam misalignment effects, highlighting the importance of accurate statistical modeling for THz links at the envelope level. Similarly,~\cite{papasotiriou2023outdoor} utilized the MG distribution to model outdoor THz measurement data at the envelope level, in addition to several other studies~\cite{9368251,c1d8db88-e5a9-3924-91ac-948eb603e81a,7430127}. Diversity techniques remain indispensable to improve the robustness and reliability of wireless links~\cite{tse2005fundamentals}. This is particularly true for THz communication systems, where severe path loss, molecular absorption, and small-scale fading significantly degrade link performance~\cite{sarieddeen2020overview}. Exploiting spatial, frequency, and polarization diversity~\cite{tse2005fundamentals,simon2001digital} can effectively mitigate these impairments and counteract the limited multipath richness typical of high-frequency channels. 

\subsection{Related Works}
Several studies have investigated the statistical modeling and performance analysis of wireless communication systems under various fading environments~\cite{10003243,9250676,10756586}. However, most existing approaches either overlook THz channel characteristics such as molecular absorption and frequency-dependent path loss~\cite{sarieddeen2020overview}, or rely on experimentally measured THz models~\cite{papasotiriou2021experimentally,papasotiriou2023outdoor} that are difficult to incorporate into tractable analytical formulations.
In~\cite{10003243}, new exact and computationally efficient expressions were derived for the PDF and CDF of the sum of $\alpha$--$\mu$ random variables (RVs), enabling precise analysis of diversity combining receivers. The work in~\cite{9250676} extended this direction by proposing two approximate closed-form solutions for the sum of i.n.i.d. $\alpha$--$\mu$ variates, and further analyzed the outage probability, bit error rate, and capacity of MIMO systems employing orthogonal space–time block coding (OSTBC) under $\alpha$--$\mu$ fading. More recently,~\cite{10756586} presented a unified analytical framework for modeling the sum of i.n.i.d. Gaussian-based perturbation fading channels, where a MG representation was used to model the SNR power. In contrast, our work here utilizes the MG model to characterize the small-scale fading behavior of outdoor THz channels at the envelope level as suggested in~~\cite{papasotiriou2023outdoor,9368251,c1d8db88-e5a9-3924-91ac-948eb603e81a,7430127}. Despite prior  advances, the existing diversity studies still overlook key THz channel characteristics such as frequency-dependent path loss and measured propagation effects that are essential for practical scenarios. Moreover, while the exact methods in~\cite{10003243,9250676} offer valuable theoretical insights, they often involve complex integrals and special functions that limit their intuitive interpretation and practical applicability in THz system analysis incorporating molecular absorption and frequency-dependent path loss.

\subsection{Contribution}
In this paper, we study the $L$-order diversity assuming a maximum ratio combiner (MRC) receiver. We provide a detailed analysis for THz channels, where the fading is modeled as MG and two variants of the $\alpha$-$\mu$ distributions for outdoor and indoor, respectively. The main contributions of this paper are summarized as follows:
 \begin{itemize}
    \item For the indoor THz scenario, we apply a simplified approximation method~\cite{tse2005fundamentals} for evaluating the diversity order under THz channel conditions that account for molecular absorption and path loss. 
    The present method yields results consistent with~\cite{10003243,9250676}, while providing a more mathematically tractable and intuitive derivation. Moreover, for the second $\alpha$-$\mu$ model~\cite{9250676}, we present an alternative analytical framework that recovers the findings of~\cite{9250676} under THz conditions using both exact and approximate methods.

    \item For the outdoor scenario, to the best of the authors' knowledge, this is the first work to derive the MRC diversity exponent for both independent and identically distributed (i.i.d) and i.n.i.d MG-modeled outdoor THz links. The analysis is carried out using the moment generating function (MGF) approach.
\end{itemize}
The proposed approximation method provides a smooth and tractable alternative to exact methods, enabling easier derivation and clearer mathematical insight while maintaining high accuracy.
We show that $\mathrm{Pr}_{\mathrm{e}}(\text{SNR}) = \Theta\left(\text{SNR}^{- \kappa_2}\right) = \kappa_1 \text{SNR}^{-\kappa_2} + o(\text{SNR}^{-\kappa_2})$, where $\kappa_2 = \frac{\alpha \mu}{2} L$ in the first $\alpha$-$\mu$ model, recovering the result of~\cite{10003243}, and $\kappa_2 =\frac{\alpha}{2} \sum_{i=1}^L \mu_i$ in the second $\alpha$-$\mu$ model, recovering the result of~\cite{9250676}. For the MG scenario, we find that $\kappa_2 = \min_{\substack{(n_1,n_2,\cdots,n_L)\\[1pt] 1\leq n_\ell \leq N_\ell, 1 \leq \ell \leq L}} \sum_{\ell = 1}^L\frac{\beta_{\ell,n_\ell}}{2}, $ for i.n.i.d, and $\kappa_2 = \left(\frac{L}{2}\right)\min\limits_{1 \leq j \leq N} \beta_j$ for the i.i.d case, where $N$ is the number of mixtures. We also determine the multiplicative constant, $\kappa_1$.

\subsection{Organization and notation}

The remainder of this paper is organized as follows. Section~\ref{sec:sysmodel} introduces the system and channel models. Section~\ref{sec:proF} presents the problem formulation. Section~\ref{sec:Indoor_ana} provides the indoor THz diversity analysis, while Section~\ref{sec:outdoor_ana} discusses the outdoor THz analysis. Section~\ref{sec:simulation} reports the simulation results, and concluding remarks are given in Section~\ref{sec:conclusion}, followed by the appendices.  

We adopt the following notation throughout the paper. Non-bold lower and upper case letters ($a, A$) denote scalars, bold lower case letters ($\mbf{a}$, $\mbf{b}$, $\cdots$) denote vectors, and bold upper case letters ($\mbf{A}$, $\mbf{B}$, $\cdots$) denote matrices. The superscripts ${(\cdot)}^\Tpow$, ${(\cdot)}^\Strpow$, ${(\cdot)}^\Hpow$, and ${(\cdot)}^\Invpow$ represent the transpose, conjugate, hermitian, and inverse operators, respectively. The $\abs{\cdot}$ operator can represent the absolute value of $a$ and the absolute value of each entry of $\mbf{a}$; $\norm{\mbf{a}}$ is the Euclidean norm of $\mbf{a}$. $\mathrm{Tr}(\cdot)$ represents the trace operator. $j=\sqrt{-1}$ denotes the imaginary unit. $\mathcal{CN}(0,\sigma^2)$ denotes a complex circularly symmetric Gaussian random variable of zero mean and variance $\sigma^2$\cite{11078009}. $\mathbb{E}[\cdot]$ is the expectation operator and $\mathrm{Pr}(\cdot)$ is the probability operator. $\Gamma(.)$ denotes the Gamma function, and \(B(x,y)\) is the Beta function $B(x,y) = \int_0^1 t^{x - 1} (1 - t)^{y - 1} dt =\frac{\Gamma(x) \Gamma(y)}{\Gamma(x+y)}$.
$Q(x)=\frac{1}{\sqrt{2 \pi}} \int_x^{\infty} e^{-\frac{u^2}{2}} d u$ is the $Q\mhyphen$function and ${Q}(x) \!=\! \frac{1}{2} \mathrm{erfc}(\frac{x}{\sqrt{2}})$ where $\text{erfc}$ is the complementary error function. $G_{p, q}^{m, n}\left[z \left\lvert \begin{smallmatrix}a_1, \ldots, a_p \\ b_1, \ldots, b_q\end{smallmatrix} \right.\right]$ represents the Meijer G-function \cite[eq. (9.301)]{prudnikov1986integrals} while $H_{p, q}^{m, n}\left[z \left\lvert \begin{smallmatrix}(a_1, b_1), \ldots, (a_p, b_p) \\ (c_1, d_1), \ldots, (c_p, d_p)\end{smallmatrix} \right.\right]$ represents the Fox H-function \cite[eq. (1.1.1)]{kilbas2004h}. We adopt the following conventions: for two asymptotically positive functions, $f(x) = o \left(g(x)\right)$ if for every $c > 0$, $f(x) \leq c g(x)$ for $x$ large enough. We write $f(x) = O\left(g(x)\right)$ if there exists $c > 0$ such that $f(x) \leq c g(x)$ for $x$ large enough.

\section{System and Channel Models}

\label{sec:sysmodel}
We consider a single-input, multiple-output (SIMO) single-carrier THz link, where the channel is modeled using a THz model propagation. Specifically, the model incorporates major physical effects observed in recent studies, including molecular absorption, frequency-selective path loss, and small-scale fading~\cite{Jemaa2024Performance,11072420,8610080,9714471}. Such modeling aligns with several recent works that adopt THz channel characterizations for THz channel scenarios. The complex baseband equivalent received signal for $L \geq 1$ receiver antennas is expressed as:
\begin{align}
\label{eq:sys_model}
    \mathbf{r} = \mathbf{h}\,x + \mathbf{n}, 
\end{align}
where \( \mathbf{\mathbf{r}} = [r_1, \ldots, r_L]^T \) is the received vector, $x$ is the modulated transmitted symbol, \( \mathbf{n} = [n_1, \ldots, n_L]^T \) is an additive white Gaussian noise with power $N_{0}$ (i.e. $n_i \sim \mathcal{CN}(0, N_0)$), and \(\mathbf{\mathbf{h}} = [h_1,\ldots, h_L]^T \) represents the overall channel fading vector with independent entries which is assumed independent of {\bf n}. Let $\nu=\sqrt{P_t G_t G_r} h_p$, then, each $h_i$, $1 \leq i \leq L$, is defined as $h=\nu h_f$, where $h_f$ is the complex small-scale fading and $h_p$ represents the THz-band free-space path loss amplitude factor, corresponding to the square-root of the received power attenuation, consisting of both spreading and molecular absorption losses, expressed as\cite{9591285} \(h_p =(c/4 \pi f d)^{\frac{\varrho}{2}} e^{-K_{\text{abs}} d/2 },\) where $c$ is the speed of light, $f$ is the operating frequency, $d$ is the communication distance, and $K_{\text{abs}}$ is the molecular absorption coefficient (more details in \cite{9591285}). In measurement-based sub-THz/THz works \cite{papasotiriou2021experimentally,9591285}, the path loss exponent, $\varrho$, is best-fit to 2. \( P_t \) denotes the transmit power, and \( G_t \) and \( G_r \) represent the gains of the transmit and receive antennas, respectively.
For indoor THz channels, we use the $\alpha$-$\mu$-distribution to represent the magnitude of $h_f$, where its PDF is expressed as~\cite{Magableh2009}
\begin{align}
\label{eq:pdff}
    f_{|h_f|}(y) = \frac{\alpha \mu^\mu y^{\alpha \mu - 1}}{\hat{Z}^{\alpha \mu} \Gamma(\mu)}  \exp\left(-\mu y^\alpha/\hat{Z}^{\alpha}\right), 
\end{align}
where $\alpha\!>\! 0$ is a fading parameter, $\mu$ is the normalized variance of the fading channel, and $\hat{Z} = \sqrt[\alpha]{\mathbb{E}(|h_f|^\alpha)}$ is the $\alpha$ root mean value of the fading channel. Using~\eqref{eq:pdff}, the PDF of $h = \nu h_f$ can thus be expressed as
\begin{align}
    f_{|h|}(y)&=\frac{1}{|\nu|}f_{|h_f|}\left(\frac{y}{\nu}\right)=\!\frac{\alpha \mu^\mu y^{\alpha \mu - 1}}{(\hat{Z}\nu)^{\alpha \mu} \Gamma(\mu)}  \exp\left(-\mu y^\alpha/(\nu\hat{Z})^{\alpha}\right),
    \label{PDF_|h|_AM}
\end{align} 
which gives
\begin{align}
    f_{|h|^2}(y) = \frac{\alpha \mu^\mu}{2 (\hat{Z}\nu)^{\alpha \mu} \Gamma(\mu)} y^{\frac{\alpha \mu }{2}-1} \exp\left(-\mu y^{\alpha/2} / (\nu \hat{Z})^\alpha \right).\label{PDF_|h|2_AM}
\end{align}
Another parameterized form of the $\alpha$–$\mu$ distribution is presented in~\cite{9250676}, as
\begin{equation}
    f_{|h_f|}(x) = \frac{\alpha \beta^{\alpha \mu}}{\bar{x}^{\alpha \mu} \Gamma(\mu)} x^{\alpha \mu - 1} \exp \left( - \left( \beta \frac{x}{\bar{x}} \right)^{\alpha} \right),
    \label{eq:alpha_generic}
\end{equation}
where $\bar{x} = \mathbb{E}\left\{\abs{h_f}\right\}$ is the average of $\abs{h_f}$ and $\beta = \frac{\Gamma(\mu + \frac{1}{\alpha})}{\Gamma(\mu)}$.
By applying a similar derivation to that used for \eqref{eq:pdff}, the PDF of \( |h|^2 \) based on \eqref{eq:alpha_generic} is obtained as
\begin{align}
    f_{|h|^2}(y)&=\!\frac{\alpha \beta^{\alpha \mu}}{2(\bar{x}\nu)^{\alpha \mu} \Gamma(\mu)} y^{\frac{\alpha \mu}{2} - 1} \exp \left( - \left( \beta \frac{\sqrt{y}}{\bar{x}\nu} \right)^{\alpha} \right). \label{PDF_|h|2_AM_generic}
\end{align}
The two PDF forms of the $\alpha$-$\mu$ distribution enable different analytical approaches and use cases, as detailed in the following sections.

For the outdoor THz model, the small-scale fading is modeled as a MG distribution~\cite{papasotiriou2023outdoor,karakoca2023measurement} whose PDF is expressed  as~\cite{papasotiriou2023outdoor}
\begin{equation}
\label{MGdist}
    f_{|h_f|}(y)\!=\!\sum_{i=1}^N \!w_i \frac{\zeta_i^{\beta_i} y^{\beta_i-1} e^{-\zeta_i y}}{\Gamma\left(\beta_i\right)}\!=\!\sum_{i=1}^N\! \alpha _i y^{\beta _i -1} e^{-\zeta _i y}, \ y\ge 0,
\end{equation}
where $N$ is the number of gamma components, and $\zeta _i$, $\beta _i$ and $w_i$ denote the scale, shape, and weight of the $\nth{i}$ component where $\sum_{i=1}^N w_i \!=\! 1$. We define $\alpha_i\!\triangleq\!w_i\zeta_i^{\beta_i}/\Gamma\left(\beta_i\right)$ for ease of notation. The PDF of $|h|$ and $|h|^2$ are given by
\begin{align}
    f_{|h|}(y) &= \sum_{i=1}^{N} \frac{\alpha_i}{\nu^{\beta_i}} y^{\beta_i - 1} e^{-\frac{\zeta_i}{\nu} y} \label{PDF_|h|_MG}\\
    f_{|h|^2}(y) &= \sum_{i=1}^{N} \frac{\alpha_i}{2 \nu^{\beta_i}} y^{\frac{\beta_i}{2}-1} e^{-\frac{\zeta_i}{\nu} \sqrt{y}}\label{PDF_|h|2_MG}
\end{align}

\section{Problem Formulation}
\label{sec:proF}
We consider an MRC receiver with perfect channel state information (CSI). The received signal is expressed as
\cite{tse2005fundamentals}
\begin{eqnarray}
    \tilde{r} = \frac{\mathbf{h}^*}{\|\mathbf{h}\|}\mathbf{r} &=& \frac{\mathbf{h}^*}{\|\mathbf{h}\|}\left(\mathbf{h} \, x + \mathbf{n}\right) = \|\mathbf{h}\|\,x + w,
\end{eqnarray}
where \(w = \frac{\mathbf{h}^*}{\|\mathbf{h}\|}\mathbf{n}  \sim \mathcal{CN}(0, N_0)\).
We define the instantaneous signal-to-noise ratio (SNR) at the receiver, relative to the channel power, as
\begin{equation}
\gamma= \frac{E_{\mathrm{s}} \norm{\mathbf{h}}^2 }{N_{0}} =\Upsilon\|\mathbf{h}\|^2,  \quad \norm{\mathbf{h}}^2=\sum_{j=1}^L \abs{h_j}^2,\label{SNR}
\end{equation}
where $\Upsilon=\frac{E_{\mathrm{s}}}{ N_{0}}$,  $E_{\mathrm{s}}$ is the energy per symbol, and each $\abs{h_j}^2$ is distributed according to~\eqref{PDF_|h|2_AM} or~\eqref{PDF_|h|2_AM_generic}, or~\eqref{PDF_|h|2_MG}.
To analyze the diversity gain, we derive expressions for the bit error probability in the form $\mathrm{Pr}_{\mathrm{e}} = \kappa_1 \Upsilon^{-\kappa_2} + o(\Upsilon^{-\kappa_2})$ in the high SNR regime (high $\Upsilon$), where $\kappa_1, \kappa_2 > 0$. To this end, we consider two approaches. 
\begin{enumerate}
\item[1-] The first approach, proposed in~\cite{tse2005fundamentals}, consists of approximating \( \mathrm{Pr}_{\mathrm{e}} \) at high $\Upsilon$ as
\begin{equation}
\mathrm{Pr}_{\mathrm{e}} \approx \mathrm{Pr}\left( \|\mathbf{h}\|^2 \leq \frac{1}{\Upsilon} \right)=\int_0^\frac{1}{\Upsilon} f_{\norm{\mathbf{h}}^2}(y)dy. \label{per_appr}
\end{equation}
At high $\Upsilon$, the PDF of \( \|\mathbf{h}\|^2 \) can be approximated for small values, which enables the derivation of approximate expressions for \( \mathrm{Pr}_{\mathrm{e}} \). We note that this method recovers the value of $\kappa_2$ but fails to determine the exact value of $\kappa_1$.
\item[2-] The second approach is to compute the exact \( \mathrm{Pr}_{\mathrm{e}} \)~\cite{simon2001digital}:
\begin{align}
    \mathrm{Pr}_{\mathrm{e}}&={\mathbb{E}}_{\|\mathbf{h}\|^2}\left[Q\left( \sqrt{\Upsilon\norm{\mathbf{h}}^2}\right)\right] \notag\\&=\int_0^{\infty} Q( \sqrt{\Upsilon x}) f_{\norm{\mathbf{h}}^2}(x) d x. \label{per_ext}
\end{align}
This method fully characterizes \( \mathrm{Pr}_{\mathrm{e}} \) as a function of $\Upsilon$ and determines both $\kappa_1$ and $\kappa_2$. 
\end{enumerate}
We note that both methods require knowledge of the PDF of $\|\mathbf{h}\|^2$ as implied by~\eqref{per_appr} and~\eqref{per_ext}.

\section{Indoor Diversity Analysis}
\label{sec:Indoor_ana}
In this section, we provide expressions of the probability of error, \(\mathrm{Pr}_{\mathrm{e}}\), using both methods as outlined in Section~\ref{sec:proF}. The analysis is carried out for the THz indoor channel models given by~\eqref{PDF_|h|2_AM} and \eqref{PDF_|h|2_AM_generic}.

\subsection{The First \texorpdfstring{$\alpha$-$\mu$}{alpha-mu} Model}

We consider i.i.d. $\alpha$-$\mu$ fading channels, i.e., all branches share identical fading parameters $(\alpha, \mu, \hat{Z})$. The diversity analysis for the $\alpha$-$\mu$ indoor channel model~\eqref{PDF_|h|_AM} is presented in~\cite{10003243} where it is found 
that 
\begin{equation}
\mathrm{Pr}_{\mathrm{e}\Upsilon\rightarrow\infty} \!= \!\!\left( \frac{\bar{\alpha} \mu^{\mu}\Gamma(\bar{\alpha} \mu)}{\Gamma(\mu)\bar{Z}^{\bar{\alpha} \mu}} \right)^L \frac{2^{-\varphi_0}\Gamma(\varphi_0+\frac{1}{2})}{\Gamma(\varphi_0+1)\sqrt{\pi}\, } \Upsilon^{-\varphi_0} + o\left(\Upsilon^{-\varphi_0}\right)\!, \label{Per_m2_sp}
\end{equation}
where $\bar{\alpha} \triangleq \frac{\alpha}{2}$ and $\bar{Z} \triangleq (\hat{Z}\nu)^2$. Equation~\eqref{Per_m2_sp} implies that \(\kappa_2 = \varphi_0 \triangleq \frac{\alpha \mu}{2} L \), highlighting its dependence on the parameters \(\alpha\), \(\mu\), and \(L\).

In this section, we recover the result of~\cite{10003243} using a partial characterization method.
The exact PDF of the sum of i.i.d. \(\alpha\)–\(\mu\) distributed random variables is derived in~\cite{10003243}:
\begin{equation}
    f_{\norm{\mathbf{h}}^2}(y) = \left( \frac{\bar{\alpha} \mu^{\mu}}{\Gamma(\mu)\bar{Z}^{\bar{\alpha} \mu}} \right)^L \sum_{i=0}^\infty \frac{\delta_i y^{i\bar{\alpha} + \bar{\alpha} \mu L - 1}}{\Gamma(i \bar{\alpha} + L \mu \bar{\alpha})},\label{norm_h_sp}
\end{equation}
 where the coefficients \(\delta_i\) are determined as  
\begin{equation}
\delta_i = 
\left\{
\begin{aligned}
&[\Gamma(\bar{\alpha} \mu)]^L, \quad i = 0 \\[1ex]
& \sum_{\ell=1}^{i} \frac{\delta_{i - \ell} (\ell L + \ell - i) 
\Gamma\big(\bar{\alpha}(\ell + \mu)\big) 
\left(-\mu \left(\tfrac{1}{\bar{Z}}\right)^{\bar{\alpha}}\right)^{\ell}}{i\, \Gamma(\bar{\alpha} \mu) \ell!}, 
\quad i \geq 1.
\end{aligned}
\right.
\label{eq:delta_i}
\end{equation}

As proposed in~\cite{tse2005fundamentals}, finding the exponent $\kappa_2$ requires approximating the PDF, \( f_{\|\mathbf{\mathbf{h}}\|^2}(y) \), for small values of \( y \). For such small values $y$ in~\eqref{norm_h_sp}, the term with the smallest exponent of $y$ dominates the series. This corresponds to the $i=0$ term in the summation. Hence, the first order approximation becomes:
\begin{equation}
    f_{\norm{\mathbf{h}}^2}(y) = \left( \frac{\bar{\alpha} \mu^{\mu}\Gamma(\bar{\alpha} \mu)}{\Gamma(\mu)\bar{Z}^{\bar{\alpha} \mu}} \right)^L \frac{ y^{\bar{\alpha} \mu L - 1}}{\Gamma(L \mu \bar{\alpha})} + o\left(y^{\bar{\alpha} \mu L - 1}\right). \label{temp2}
\end{equation}
Evaluating~\eqref{per_appr} for large $\Upsilon$ results in
\begin{align}
      \mathrm{Pr}\left( \|\mathbf{h}\|^2 \leq \frac{1}{\Upsilon} \right)= &\int_0^\frac{1}{\Upsilon} f_{\norm{\mathbf{h}}^2}(y)dy\notag \\
    = & \left( \frac{\bar{\alpha} \mu^{\mu}\Gamma(\bar{\alpha} \mu)}{\Gamma(\mu)\bar{Z}^{\bar{\alpha} \mu}} \right)^L \!\!\!\frac{ \Upsilon^{-\bar{\alpha} \mu L}}{\Gamma(\bar{\alpha} \mu L+1)} + o\left(\!\!\Upsilon^{-\bar{\alpha} \mu L}\!\!\right),\label{Per_m1_sp}
\end{align}
where the interchange in the order of the integral and the small-o notation can be justified by the  Lebesgue's dominated convergence theorem (DCT). Thus, the diversity exponent is given by \( \kappa_2 = \bar{\alpha} \mu L = \frac{\alpha \mu}{2} L \).

\subsection{The Second \texorpdfstring{$\alpha$-$\mu$}{alpha-mu} Model}
In contrast to the previous subsection, here we consider the more general i.n.i.d.\ case, where the fading parameters $\mu_i$ and $\hat{Z}_i$ may vary across branches. We consider the THz indoor channel using the $\alpha$-$\mu$ fading model, whose PDF is given in~\eqref{PDF_|h|2_AM_generic}. To derive the PDF of \(\|\mathbf{h}\|^2\) in this case, we use the result of~\cite{jemaa2025performancecomplexityanalysisterahertzband} where it has been shown that
\begin{equation}
    f_{\norm{\mathbf{h}}^2}(y) \approx \sum_{m=1}^{\Psi} \frac{c_{m} \bar{\alpha} \bar{\beta}^{\bar{\alpha} \bar{\mu}} y^{\bar{\alpha} \bar{\mu} - 1}}{(\omega_{m} \bar{z})^{\bar{\alpha} \bar{\mu}} \Gamma(\bar{\mu})} \,\exp{\left(-\left(\bar{\beta} \frac{y}{\omega_{m} \bar{z}}\right)^{\bar{\alpha}}\right)},\label{pdf_h_2_gen}
\end{equation}
where
\[
\bar{\alpha} = \frac{\alpha}{2}, \quad \bar{\mu} = \sum_{i=1}^{L} \mu_i, \quad \bar{\beta} = \frac{\Gamma\left(\bar{\mu} + \frac{1}{\bar{\alpha}}\right)}{\Gamma(\bar{\mu})}, \quad \bar{z} = \nu^2 \sum_{i=1}^L \bar{x}_i^2.
\]
For
integer $\Psi \geq 2$ controls the approximation accuracy of the sum of $\alpha$-$\mu$ random variables; as $\Psi \rightarrow \infty$, \eqref{pdf_h_2_gen} becomes exact. For any arbitrary integer, \(\Psi\), the parameters \(c_m\) and \(\omega_m\) are determined by solving the system of linear equations:
\begin{equation}
    \sum_{m=1}^{\Psi} c_m \omega_m^n = \frac{\mathbb{E}\left[Z^n\right]}{\mathbb{E}^n\left[Z\right]} \xi^{(n)}, \quad n = 0, 1, 2, \ldots, 2\Psi-2,
\end{equation}
\begin{equation}
    \sum_{m=1}^{\Psi} \frac{c_m}{\omega_m^{\bar{\alpha} \bar{\mu}}} = \bar{\alpha}^{L-1} \frac{\bar{z}^{\bar{\alpha} \bar{\mu}} \Gamma(\bar{\mu})}{\bar{\beta}^{\bar{\alpha} \bar{\mu}} \Gamma(\bar{\alpha} \bar{\mu})} \times \prod_{i=1}^{L} \frac{\beta_i^{\bar{\alpha} \mu_i} \Gamma(\bar{\alpha} \mu_i)}{\bar{y}_i^{\bar{\alpha} \mu_i} \Gamma(\mu_i)}.
\end{equation}
Here, the random variable $Z = \|\mathbf{h}\|^2$ represents the sum $Z = Y_1 + Y_2 + \dots + Y_L$, where each $Y_i$ is distributed as $|h|^2$ according to the $\alpha$-$\mu$ PDF defined in~\eqref{PDF_|h|2_AM_generic}, and \(\xi^{(n)}\) is defined as \(\xi^{(n)} = \frac{\Gamma(\bar{\mu} + \frac{n}{\bar{\alpha}}) \Gamma^{n-1}(\bar{\mu})}{\Gamma^n (\bar{\mu} + \frac{1}{\bar{\alpha}})}.
\)
To derive \(\mathbb{E}[Z^n]\), we use the following formula, which relies on the individual terms of the summation in the random variable \(Z\)~\cite{jemaa2025performancecomplexityanalysisterahertzband}:
{
\begin{align}
E(Z^n) &= \sum_{n_1=0}^{n} \sum_{n_2=0}^{n_1} \cdots \sum_{n_{L-1}=0}^{n_{L-2}} 
\binom{n}{n_1} \binom{n_1}{n_2} \cdots \binom{n_{L-2}}{n_{L-1}} \nonumber \\
&\quad \times E\left(Y_1^{n - n_1}\right) E\left(Y_2^{n_1 - n_2}\right) 
\cdots E\left(Y_L^{n_{L-1}}\right).
\end{align}
}
We proceed next to finding $\kappa_1$ and $\kappa_2$ for the THz model in~\eqref{PDF_|h|2_AM_generic}, using~\eqref{pdf_h_2_gen}.

\subsubsection{Full Diversity Characterization}
Starting with~\eqref{per_ext} and using~\eqref{pdf_h_2_gen}, we obtain

\begin{align}
& \mathrm{Pr}_{\mathrm{e}}=\sum_{m=1}^{\Psi} \frac{\Lambda_m}{2} \int_0^{\infty} y^{\bar{\alpha} \bar{\mu} - 1} \text{erfc}\left(\frac{ \sqrt{\Upsilon y}}{\sqrt{2}}\right)\notag\\&\hspace{35mm}\times\exp{\left(-\left(\bar{\beta} \frac{y}{\omega_{m} \bar{z}}\right)^{\bar{\alpha}}\right)}dy,\label{temp23}
\end{align} 
where \[\Lambda_m \triangleq \frac{c_{m} \bar{\alpha} \bar{\beta}^{\bar{\alpha} \bar{\mu}}}{(\omega_{m} \bar{z})^{\bar{\alpha} \bar{\mu}} \Gamma(\bar{\mu})}.\] The integral has been evaluated in prior work following this exact approach~\cite[Eq.~(39)]{jemaa2025performancecomplexityanalysisterahertzband},~\cite{WolframMeijerGIntegration}, which provide a complete derivation and its reduction to Fox--$H$ forms. The integral in~\eqref{temp23} is absolutely convergent under the following conditions, satisfied for all parameter values considered in this work: (i) $\bar{\alpha}>0$ and $\bar{\beta}>0$, ensuring that $\exp[-(\bar{\beta} y/(\omega_m\bar{z}))^{\bar{\alpha}}]$ decays rapidly as $y \to \infty$; (ii) $\Upsilon>0$, so that $\text{erfc}(\sqrt{\Upsilon y}/\sqrt{2})$ remains bounded near $y=0$ and decays exponentially for large $y$; (iii) $\omega_m\bar{z}>0$, guaranteeing a positive and real exponential argument; and (iv) $\bar{\alpha}\bar{\mu}>0$, ensuring integrability at the origin since the integrand behaves as $y^{\bar{\alpha}\bar{\mu}-1}$ for $y \to 0$. Under these conditions, the combined effect of the $\text{erfc}$ tail and the stretched-exponential term ensures exponential or faster decay as $y \to \infty$, confirming the convergence of the integral.
As a result, we obtain:
\begin{align}
   \mathrm{Pr}_{\mathrm{e}} &= \sum_{m=1}^{\Psi} \frac{\Lambda_m}{2\sqrt{\pi}} \left( \frac{\Upsilon}{2}\right)^{-\bar{\alpha} \bar{\mu}} \notag \\ &\times H_{2,2}^{1,2} \left[  \left( \frac{2\bar{\beta}}{\Upsilon\omega_{m} \bar{z}}\right)^{\bar{\alpha}}\Bigg| \begin{array}{c} (1-\bar{\alpha} \bar{\mu}, \bar{\alpha})(\frac{1}{2}-\bar{\alpha} \bar{\mu}, \bar{\alpha}) \\ (0,1)\, (-\bar{\alpha}\bar{\mu}, \bar{\alpha}\bar{\mu}) \end{array} \right].\label{per_m2_gen}
\end{align}
For \(\Upsilon \to \infty\), the term \( t = \left( \frac{2\bar{\beta}}{\Upsilon\omega_{m} \bar{z}}\right)^{\bar{\alpha}} \) in~\eqref{per_m2_gen} approaches zero. In \cite[Appendix F]{Jemaa2024Performance}~\cite[Eq. 3.6]{kilbas1998hfunction}, the authors conducted an asymptotic analysis of the Fox H-function \( H_{p,q}^{s,r}(t|.) \) in the regime where \( t \to 0 \) to obtain:
\begin{align}
H_{p,q}^{s,r}&\Biggl[
z \;\Bigg|
\begin{matrix}
(a_1,\alpha_1), \dots, (a_p,\alpha_p) \\
(b_1,\beta_1), \dots, (b_q,\beta_q)
\end{matrix}
\Biggr]\notag\\&= \sum_{j=1}^s \left[ h_j^* z^{b_j/\beta_j} + O\left(z^{(b_j+1)/\beta_j}\right)\right] \quad (z \to 0), 
\label{eq:appfox}
\end{align}
\begin{equation}
    h_j^* = \frac{\prod_{i=1,i\neq j}^{s} \Gamma\left(b_i - \frac{b_j\beta_i}{\beta_j}\right) \prod_{i=1}^r \Gamma\left(1 - a_i + \frac{b_j\alpha_i}{\beta_j}\right)}{\beta_j \prod_{i=r+1}^p \Gamma\left(a_i - \frac{b_j\alpha_i}{\beta_j}\right) \prod_{i=s+1}^q \Gamma\left(1 - b_i + \frac{b_j\beta_i}{\beta_j}\right)}.
\end{equation}
Using the asymptotic expression~\eqref{eq:appfox}, for \(\Upsilon \to \infty\), equation~\eqref{per_m2_gen} becomes  
\begin{equation} \mathrm{Pr}_{\mathrm{e}\Upsilon\rightarrow\infty}\approx \frac{2^{\bar{\alpha}\bar{\mu} - 1}}{\sqrt{\pi}} \left(\sum_{m=1}^{\Psi}\Lambda_m\right) \Upsilon^{-\bar{\alpha}\bar{\mu}}\left[h_1^*  +O\left( \frac{2\bar{\beta}}{\Upsilon\omega_{m} \bar{z}}\right)\right], \label{Per_m2_2}
\end{equation}
where:
\begin{equation}
    h_1^*=\frac{\Gamma\left(\bar{\alpha}\bar{\mu}\right)\, \Gamma\left(\tfrac{1}{2}+\bar{\alpha}\bar{\mu}\right)}{\Gamma\left(1+\bar{\alpha}\bar{\mu}\right)}\notag.
\end{equation}
As observed, the diversity exponent is given by \( \kappa_2 = \bar{\alpha} \bar{\mu} = \frac{\alpha}{2} \sum_{j=1}^{L} \mu_j \). This quantifies the rate at which the probability of error decreases with increasing SNR, where higher values of $\bar{\alpha} \bar{\mu}$ imply higher diversity gains. In addition, we find $\kappa_1 =   \frac{2^{\bar{\alpha}\bar{\mu} - 1}}{\sqrt{\pi}} \left(\sum_{m=1}^{\Psi}\Lambda_m\right) h_1^* $. Our results in~\eqref{Per_m2_2} are consistent with those reached in~\cite{9250676} as both methods yield the same diversity exponent.

\subsubsection{Partial Diversity Characterization}
We recover $\kappa_2$ using the approximation method. 
When \( y \) is sufficiently small, we use the approximation $\exp(y) \approx 1 + o(y)$, where~\eqref{pdf_h_2_gen} boils down to
\begin{equation}
    f_{\norm{\mathbf{h}}^2}(y) \approx \sum_{m=1}^{\Psi} \Lambda_m y^{\bar{\alpha} \bar{\mu} - 1} + o\left(y^{\bar{\alpha} \bar{\mu} - 1}\right),\label{pdf_h_2_gen_app}
\end{equation}
this implies that, for large values of $\Upsilon$, 
\begin{align}
    \mathrm{Pr}\left( \|\mathbf{h}\|^2 \leq \frac{1}{\Upsilon} \right) = & \sum_{m=1}^{\Psi} \Lambda_m  \int_0^\frac{1}{\Upsilon} y^{\bar{\alpha} \bar{\mu} - 1} dy + \int_0^\frac{1}{\Upsilon} o\left(y^{\bar{\alpha} \bar{\mu} - 1}\right) dy
 \notag\\
  \approx & \left(\sum_{m=1}^{\Psi} \Lambda_m\right)  \frac{\Upsilon^{-\bar{\alpha} \bar{\mu}}}{\bar{\alpha} \bar{\mu}}  + o\left(\Upsilon^{-\bar{\alpha} \bar{\mu}}\right),\label{per_m1_ge}
\end{align}
Equation~\eqref{per_m1_ge} implies that the diversity exponent is equal to $\kappa_2 = \bar{\alpha} \bar{\mu} = \frac{\alpha}{2}\sum_{j=1}^L\mu_j$, which is consistent with the result of the full characterization in~\eqref{Per_m2_2}.

\section{Outdoor Diversity Analysis}
\label{sec:outdoor_ana}
This section analyzes the diversity behavior of outdoor THz channels for both the i.i.d and i.n.i.d fading branches.

\subsection{I.I.D Outdoor THz Diversity Analysis}
Assuming i.i.d. fading branches, the average bit-error probability of an $L$-branch MRC receiver can be expressed in terms of the MGF as~\cite[eq. 9.12]{simon2001digital}:
\begin{equation}
    \mathrm{Pr}_{\mathrm{e}} = \frac{1}{\pi} \int_{0}^{\pi/2} 
    \left( \mathcal{M}_{\gamma'} \!\left( -\frac{g}{\sin^2 \theta} \right) \right)^{\!L} 
    d\theta,
\end{equation}
where \( g = 1 \) for binary phase-shift keying (BPSK), and \( \gamma' = \Upsilon |h|^2 \), with \( |h|^2 \) defined in~\eqref{PDF_|h|2_MG}. 
Alternatively, the Laplace transform can be employed instead of the MGF, since \( M_{\gamma'}(t) = \mathcal{L}\{ f_{\gamma'} \}(-t) \)~\cite{kobayashi2011probability}. We get
\begin{equation}
    \mathrm{Pr}_{\mathrm{e}} = \frac{1}{\pi} \int_{0}^{\pi/2} 
    \left( \mathcal{L}\{ f_{\gamma'} \} \!\left( \frac{g}{\sin^2 \theta} \right) \right)^{\!L} 
    d\theta,
    \label{SER_lap}
\end{equation}
where \( \mathcal{L}\{ f_{\gamma'} \}(\cdot) \) denotes the Laplace transform of \( f_{\gamma'}(\cdot) \). In Appendix \ref{appA}, we derive a closed-form expression of $\mathcal{L}{\{f_{\gamma'}\}}(.)$ and use it to evaluate \( \mathrm{Pr}_{\mathrm{e}} \) at high SNR to extract the corresponding diversity exponent.
Using the high SNR approximation of $\mathcal{L}{\{f_{\gamma'}\}}(.)$ given by~\eqref{approx_lap}, we apply the multinomial expansion to the resulting sum raised to the $L$-th power~\cite{gradshteyn2014table,prudnikov1986integrals} to find an approximate expression of~\eqref{SER_lap} for large $\Upsilon$:
\begin{align}
&\mathrm{Pr}_{\mathrm{e}\Upsilon\rightarrow\infty} \nonumber\\
&\approx \frac{1}{\pi} \sum_{\substack{k_1 + k_2 + \cdots + k_N = L \\ k_1, k_2, \ldots, k_N \geq 0}}\frac{L!}{k_1!\, k_2!\, \cdots\, k_N!}\Upsilon^{-\sum_{j=1}^Nk_jb_j} \notag\\
& \qquad \times \frac{\prod_{j=1}^N \left(a_j\Gamma(b_j)\right)^{k_j}}{g^{\sum_{j=1}^Nb_j k_j}}\times\int_0^{\pi/2} \left(\sin^2(\theta)\right) ^{ \sum_{j=1}^Nb_j k_j} d\theta\notag\\
&\approx  \frac{1}{\pi} \sum_{\substack{k_1 + k_2 + \cdots + k_N = L \\ k_1, k_2, \ldots, k_N \geq 0}}\frac{L!}{k_1!\, k_2!\, \cdots\, k_N!}\Upsilon^{-\sum_{j=1}^Nk_jb_j} \notag\\
&\qquad \quad \quad \times\frac{\prod_{j=1}^N \left(a_j\Gamma(b_j)\right)^{k_j}}{g^{\sum_{j=1}^Nb_j k_j}}2^{\Phi-2} \mathrm{B}\left(\frac{\Phi}{2},\frac{\Phi}{2}\right),\label{SER_MG}
\end{align} 
where $a_j = \alpha_j / (2\nu^{\beta_j})$, $b_j = \beta_j / 2$, $\Phi=2\sum_{j=1}^Nb_j k_j +1$, and $\mathrm{B}(.,.)$ represent the beta function which is given as \(\mathrm{B}\left(x,y \right) =\frac{\Gamma(y)\Gamma(x)}{\Gamma(x+y)}\), and where we used standard beta function identities~\cite[3.621.1]{gradshteyn2014table} to justify equation~\eqref{SER_MG}. This expansion enables us to express the error probability as a finite sum of terms that clearly expose the diversity order and its dependence on system parameters.
The multinomial expansion in~\eqref{SER_MG} reveals that the dominant term arises when \( k_j = L \) for the index corresponding to the smallest shape parameter, i.e., \(\min\limits_{1 \leq j \leq N} \beta_j\). Accordingly, the diversity exponent is obtained as \(\kappa_2 = \frac{L}{2} \min\limits_{1 \leq j \leq N} \beta_j.\)
Similarly, the associated multiplicative constant is expressed as
\begin{align}
    \kappa_1 = &\frac{\left(a_{min} \Gamma(\tfrac{\beta_{min}}{2})\right)^{L}}{g^{ \tfrac{L \beta_{min}}{2}}}2^{\, L \beta_{min} - 1} \notag\\&\hspace{25mm}\times\mathrm{B}\!\left(\frac{ L \beta_{min} + 1}{2}, \frac{ L \beta_{min} + 1}{2}\right),
\end{align}
where the component satisfying \( \beta_{min} \triangleq \min\limits_{1 \leq j \leq N} \beta_j \). For $L = 1$, our result exactly matches the results of~\cite{Jemaa2024Performance}, where the error probability $\mathrm{Pr}_{\mathrm{e}}$ is derived for the SISO case.

\subsection{I.n.I.D Outdoor THz Diversity Analysis}
We consider the case of i.n.i.d.\ MG fading channels, where each of the $L$ diversity branches follows a distinct MG distribution. 
Similar to the i.i.d.\ case, the average bit error probability can be expressed as~\cite[Eq.~(9.11)]{simon2001digital}
\begin{equation}
    \mathrm{Pr}_{\mathrm{e}} 
    = \frac{1}{\pi} \int_{0}^{\pi/2} 
    \prod_{\ell=1}^{L} 
    \mathcal{L}\!\left\{ f_{\gamma'_\ell} \right\}
    \!\left( \frac{g}{\sin^2 \theta} \right)
    d\theta,
    \label{SER_lap_non_iid}
\end{equation}
where $\gamma'_\ell = \Upsilon |h_\ell|^2$, and $\mathcal{L}\{f_{\gamma'_\ell}\}(s)$ denotes the Laplace transform of $\gamma'_\ell$ at the $\ell$-th branch.

Using the MG-based approximation of the Laplace transform defined in Appendix~\ref{appA}, Eq.~\eqref{approx_lap}, and substituting $s = \frac{g}{\sin^2 \theta}$, we obtain
\begin{equation}
    \mathcal{L}\{f_{\gamma'_\ell}\}\!\left(\frac{g}{\sin^2 \theta}\right)
    \approx 
    \sum_{n=1}^{N_\ell} 
    A_{\ell,n}\, g^{-\frac{\beta_{\ell,n}}{2}} (\sin \theta)^{\beta_{\ell,n}},
\end{equation}
where 
\[
A_{\ell,n} = \frac{\alpha_{\ell,n}\, \Upsilon^{-\frac{\beta_{\ell,n}}{2}}\, \Gamma\!\left(\tfrac{\beta_{\ell,n}}{2}\right)}{2\, \nu^{\beta_{\ell,n}}}.
\]

By substituting this expression into~\eqref{SER_lap_non_iid}, the average bit error probability can be approximated as
\begin{equation}
    \mathrm{Pr}_{\mathrm{e}\Upsilon\rightarrow\infty} 
    \approx 
    \frac{1}{\pi} \int_{0}^{\pi/2}
    \prod_{\ell=1}^{L}
    \left(
    \sum_{n_\ell=1}^{N_\ell} 
    A_{\ell,n_\ell}\, g^{-\frac{\beta_{\ell,n_\ell}}{2}} (\sin \theta)^{\beta_{\ell,n_\ell}}
    \right)
    d\theta.
    \label{eq:SER_nonIID_expanded}
\end{equation}

Expanding the product term and interchanging the finite sums and the integral yield
\begin{align}
    &\mathrm{Pr}_{\mathrm{e}\Upsilon\rightarrow\infty} 
    \approx\notag\\ 
    &\quad\frac{1}{\pi} 
    \sum_{n_1=1}^{N_1}\!\!\cdots\!\!\sum_{n_L=1}^{N_L}
    \left(
        \prod_{\ell=1}^{L} 
        A_{\ell,n_\ell}\, g^{-\frac{\beta_{\ell,n_\ell}}{2}}
    \right)
    \int_{0}^{\pi/2}\!
        (\sin \theta)^{\sum_{\ell=1}^{L}\beta_{\ell,n_\ell}}\, d\theta.\label{temp_niid}
\end{align}
By integrating term-by-term using~\cite[Eq.~(3.621.1)]{gradshteyn2014table}, the approximate average error probability can be expressed in closed form for the asymptotic expression~\eqref{temp_niid} as
\begin{align}
\mathrm{Pr}_{\mathrm{e}\Upsilon\rightarrow\infty} &\approx \frac{1}{\pi}
\sum_{n_1=1}^{N_1}\cdots\sum_{n_L=1}^{N_L}
\Bigg(\prod_{\ell=1}^{L} A_{\ell,n_\ell}\, g^{-\frac{\beta_{\ell,n_\ell}}{2}}\Bigg)
\nonumber\\
&\qquad\times 2^{\sum_{\ell=1}^{L}\beta_{\ell,n_\ell}-1}\,
\mathrm{B}\Big(\tfrac{\sum_{\ell=1}^{L}\beta_{\ell,n_\ell}\,+1}{2},
\tfrac{\sum_{\ell=1}^{L}\beta_{\ell,n_\ell}\,+1}{2}\Big),\nonumber
\\&\approx \frac{1}{\pi}
\sum_{n_1=1}^{N_1}\cdots\sum_{n_L=1}^{N_L}
\Bigg(\prod_{\ell=1}^{L} \frac{\alpha_{\ell,n_\ell}\, \Gamma\!\left(\tfrac{\beta_{\ell,n_\ell}}{2}\right)}{2\, \nu^{\beta_{\ell,n_\ell}}\,g^{\frac{\beta_{\ell,n_\ell}}{2}}}\Bigg)\,\Upsilon^{-\sum_{\ell=1}^{L}\frac{\beta_{\ell,n_\ell}}{2}}
\nonumber\\
&\qquad\times 2^{\sum_{\ell=1}^{L}\beta_{\ell,n_\ell}\,-\,1}\,
\mathrm{B}\Big(\tfrac{\sum_{\ell=1}^{L}\beta_{\ell,n_\ell}\,+\,1}{2},
\tfrac{\sum_{\ell=1}^{L}\beta_{\ell,n_\ell}\,+\,1}{2}\Big).
\label{eq:SER_nonIID_closedform}
\end{align}
Equation~\eqref{eq:SER_nonIID_closedform} provides a general closed-form approximation for the average bit error rate (BER) over i.n.i.d.\ MG-modeled fading channels at high SNR($\Upsilon\rightarrow\infty$). The diversity exponent is obtained as 
\begin{equation}
\kappa_2 = \min_{\substack{(n_1,n_2,\cdots,n_L)\\[1pt] 1\leq n_\ell \leq N_\ell, 1 \leq \ell \leq L}} \sum_{\ell = 1}^L\frac{\beta_{\ell,n_\ell}}{2}, 
\label{eq:attmin}
\end{equation}
where $N_\ell\in\left\{ N_1, N_2,\dots, N_L\right\}$.
Similarly, the associated multiplicative constant is expressed as
\begin{align}
    \kappa_1 &= \Bigg(\prod_{\ell=1}^{L} \frac{\alpha_{\ell,n_\ell'}\, \Gamma\!\left(\tfrac{\beta_{\ell,n_\ell'}}{2}\right)}{2\, \nu^{\beta_{\ell,n_\ell'}}\,g^{\frac{\beta_{\ell,n_\ell'}}{2}}}\Bigg)2^{\sum_{\ell=1}^{L}\beta_{\ell,n_\ell'}\,-\,1}\,
\notag\\&\qquad\qquad\times\mathrm{B}\Big(\tfrac{\sum_{\ell=1}^{L}\beta_{\ell,n_\ell'}\,+\,1}{2},
\tfrac{\sum_{\ell=1}^{L}\beta_{\ell,n_\ell'}\,+\,1}{2}\Big),
\end{align}
where $\{n_\ell'\}_{1 \leq \ell \leq L}$ denote the $L$ indices corresponding to the attained minimum in~\eqref{eq:attmin}.

It is worth mentioning that the components $\kappa_1$ and $\kappa_2$ reduce to the i.i.d.\ case derived earlier when all channel parameters are identical.

\begin{table}[!t]
\centering
\caption{{Parameter configurations for MG.}}
\resizebox{\columnwidth}{!}{%
\begin{tabular}{c c c c c c}
\hline
$N$ & $w$ & $\beta$ & $\zeta$ & \textbf{Ref.} & \textbf{Configuration} \\
\hline
\multirow{2}{*}{N=2} & 0.67540627 & 15.2709327  & 0.069986341 & \multirow{2}{*}{Ref.~\cite{papasotiriou2023outdoor}}& \multirow{2}{*}{Config. 1}  \\
 & 0.32459373 & 4.417045104 & 0.153163953 &  \\
\hline
\multirow{2}{*}{N=2} & 0.512500204 & 3.75341946  & 0.159416427 & \multirow{2}{*}{Ref.~\cite{papasotiriou2023outdoor}}& \multirow{2}{*}{Config. 2} \\
 & 0.487499796 & 22.59894871 & 0.054461913 &  \\
\hline
\multirow{3}{*}{N=3} & 0.536820771 & 11.86214023  & 0.073313171 & \multirow{3}{*}{Ref.~\cite{papasotiriou2023outdoor}}& \multirow{3}{*}{Config. 3} \\
 & 0.319060454 & 34.27672813 & 0.036198297 &  \\
 & 0.144118775 & 3.511663936 & 0.155541075 &  \\
\hline
\multirow{3}{*}{N=3} & 0.382437165 & 3.363382854 & 0.158117222 & \multirow{3}{*}{Ref.~\cite{papasotiriou2023outdoor}}& \multirow{3}{*}{Config. 4} \\
 & 0.366850269 & 20.56338721 & 0.046248736 &  \\
 & 0.250712566 & 65.00282197 & 0.021761466 &  \\
\hline
\end{tabular}%
}
\label{tab:MG}
\end{table}

\section{Simulation Results}
\label{sec:simulation}

We validate the theoretical analysis through Monte Carlo simulations configured with parameters derived from THz channel measurements~\cite{papasotiriou2021experimentally,9591285}. Unless otherwise stated, each simulated point is obtained by averaging over $5\times10^6$ independent realizations per SNR value to improve the accuracy and reliability of the results. We note that the standard errors of the Monte Carlo SER values are found to be within $[\,3.5\times10^{-6},\,7.4\times10^{-4}]$, confirming the statistical stability and accuracy of the simulation results.
We define $N_0 \!=\! k_{\mathrm{B}}TB$, where $k_{\mathrm{B}}$ is the Boltzmann constant, $B$ is the system bandwidth, and $T\!=\!\unit[300]{K} $ is the system temperature. Simulations are performed under BPSK modulation with $f\!=\!\unit[0.142]{THz}$ and $B\!=\!\unit[4]{GHz}$. The used antenna gains are $G_r \!=\! \unit[19]{dBi}$ and $G_t \!=\! \unit[0]{dBi}$ \cite{papasotiriou2023outdoor,9591285}. We adjust $P_{t}$ to vary the SNR range. For indoor small-scale fading, we evaluate the $\alpha$-$\mu$ model from~\cite{papasotiriou2021experimentally} with different parameters.

\begin{figure}[t!]
 \centering
 \includegraphics[width=0.495\textwidth]{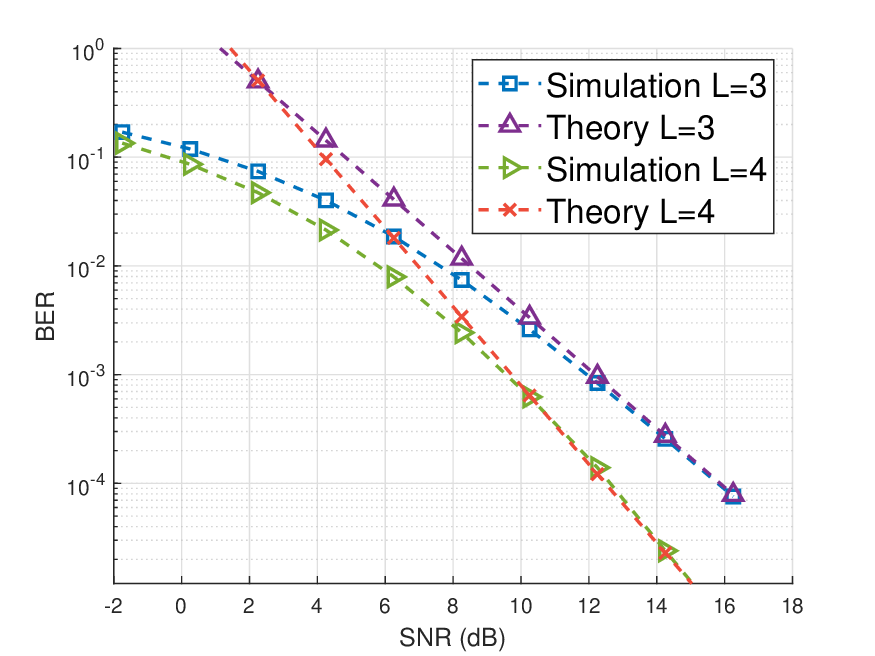}
 \caption{BER performance versus SNR under $\alpha=3.45388$ and $\mu=0.51571$ for diversity orders $L=3$ and $L=4$. Theoretical results are based on \eqref{Per_m2_sp}.}
 \label{fig-diversity_a1_sp}
\end{figure}

\begin{figure}[t!]
 \centering
 \includegraphics[width=0.495\textwidth]{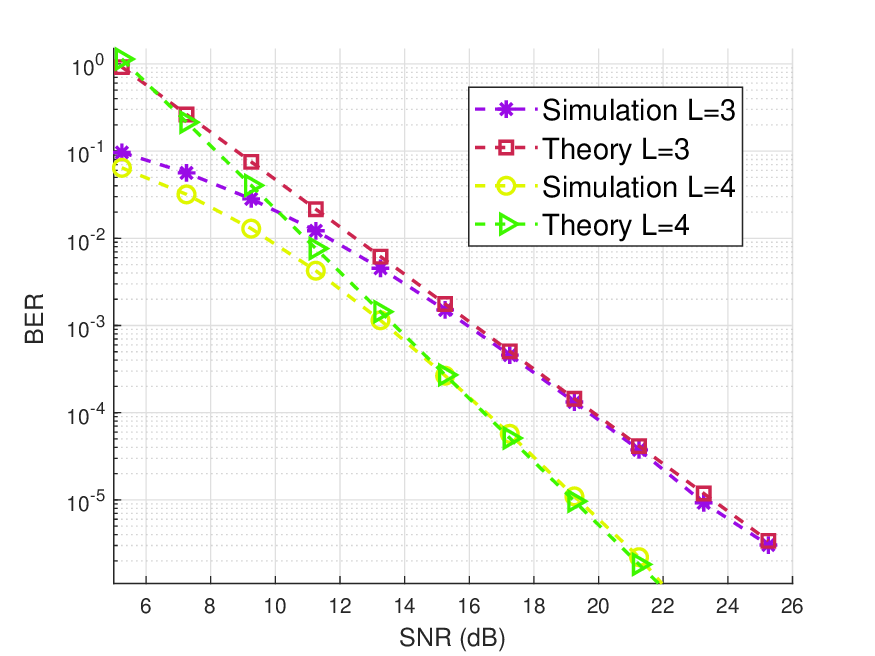}
 \caption{BER performance versus SNR under $\alpha=2.92801$ and $\mu=0.61844$ for diversity orders $L=3$ and $L=4$. Theoretical results are based on \eqref{Per_m2_sp}.}
 \label{fig-diversity_a2_sp}
\end{figure}

\begin{figure}[t!]
 \centering
 \includegraphics[width=0.495\textwidth]{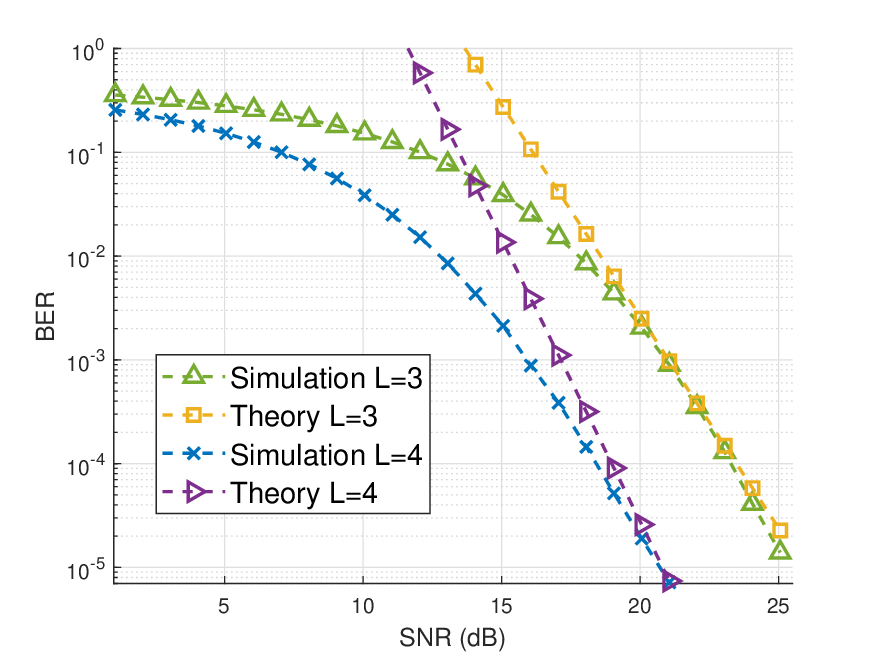}
 \caption{BER performance versus SNR for diversity orders $L=3$ and $L=4$. Theoretical results are based on \eqref{Per_m2_2}.}
 \label{fig:diversity_method_2}
\end{figure}

Figure~\ref{fig-diversity_a1_sp} depicts the BER performance for the $\alpha$-$\mu$ distribution with parameters $\alpha = 3.45388$ and $\mu = 0.51571$, as derived from experimental measurements in \cite{papasotiriou2021experimentally}. The results are shown for two diversity configurations: $L = 3$ and $L = 4$. The theoretical analysis is based on the results of the full characterization method as captured by equation~\eqref{Per_m2_sp}. The partial characterization method in~\eqref{Per_m1_sp} predicts the same diversity exponent $\kappa_2$, and its accuracy is thus captured by the plot of the exact method. The close agreement between~\eqref{Per_m2_sp} and the simulation results in the high SNR regime confirms the validity of our analysis. As anticipated, increasing the number of diversity branches $L$ leads to significant performance improvements, highlighting the effectiveness of diversity in mitigating fading effects. Fig.~\ref{fig-diversity_a2_sp} depicts the outcome of a similar setup but for a different set of channel parameters, $\alpha = 2.92801$ and $\mu = 0.61844$\cite{papasotiriou2021experimentally}. A third numerical example is presented in Fig.~\ref{fig:diversity_method_2} where the second form of the $\alpha$-$\mu$ channel given by~\eqref{PDF_|h|2_AM_generic} is considered. The model parameters are taken from~\cite{papasotiriou2021experimentally}. Once more, we observe a close match in the high SNR regime between the theoretical and simulated curves. 


\begin{figure}[t!]
    \centering
    \includegraphics[width=0.495\textwidth]{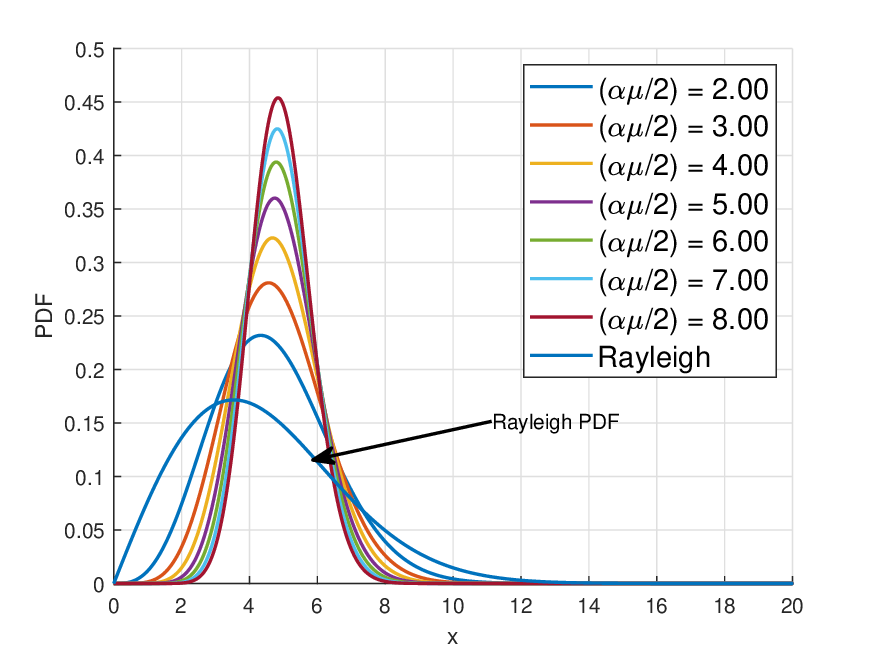}
    \caption{Probability density function of the $\alpha$-$\mu$ distribution for fixed $\alpha = 2$ and varying $\mu$ values.}
    \label{fig:PDF_fixed_alpha}
\end{figure}

\begin{figure}[t!]
    \centering
    \includegraphics[width=0.495\textwidth]{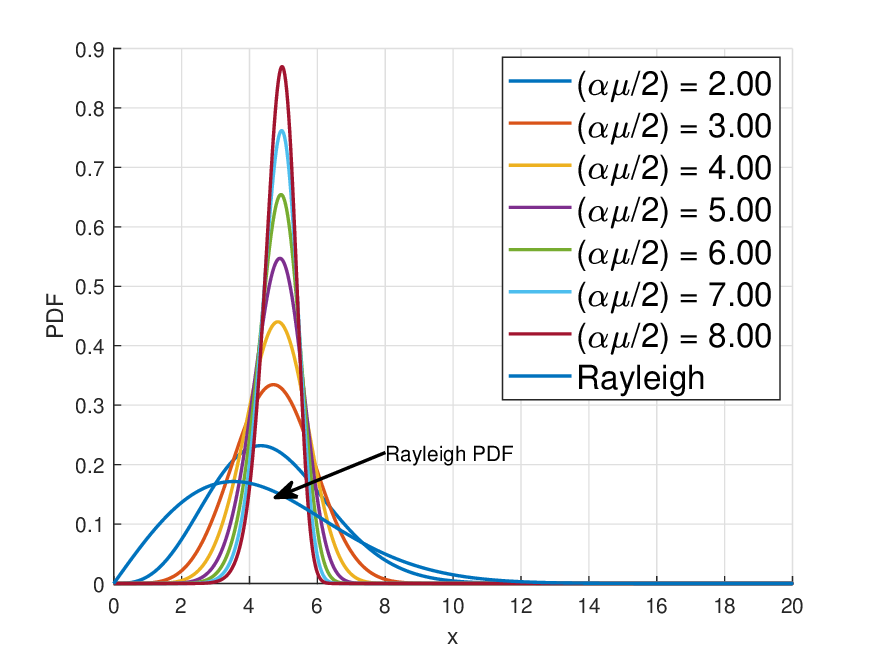}
    \caption{Probability density function of the $\alpha$-$\mu$ distribution for fixed $\mu = 2$ and varying $\alpha$ values.}
    \label{fig:PDF_fixed_mu}
\end{figure}

The decrease in \( \mathrm{Pr}_{\mathrm{e}} \) with \( \kappa_2 = \frac{\alpha \mu}{2} L \) as a function of SNR can be explained by the fact that $\alpha$-$\mu$ PDF becomes more concentrated as the product \( \alpha \mu \) increases. In Fig.~\ref{fig:PDF_fixed_alpha} and Fig.~\ref{fig:PDF_fixed_mu}, we plot the $\alpha$-$\mu$ PDF for various values of $\mu$ and $\alpha$, respectively. The figures highlight the effect of the product $\alpha \mu$ on the density function: the greater the $\alpha \mu$ product, the higher the concentration of the PDF around its LoS component (the better the channel), which is in line with the diversity exponent expression as captured by $\kappa_2$. 

\begin{figure}[t!]
     \centering
    \includegraphics[width=0.495\textwidth]{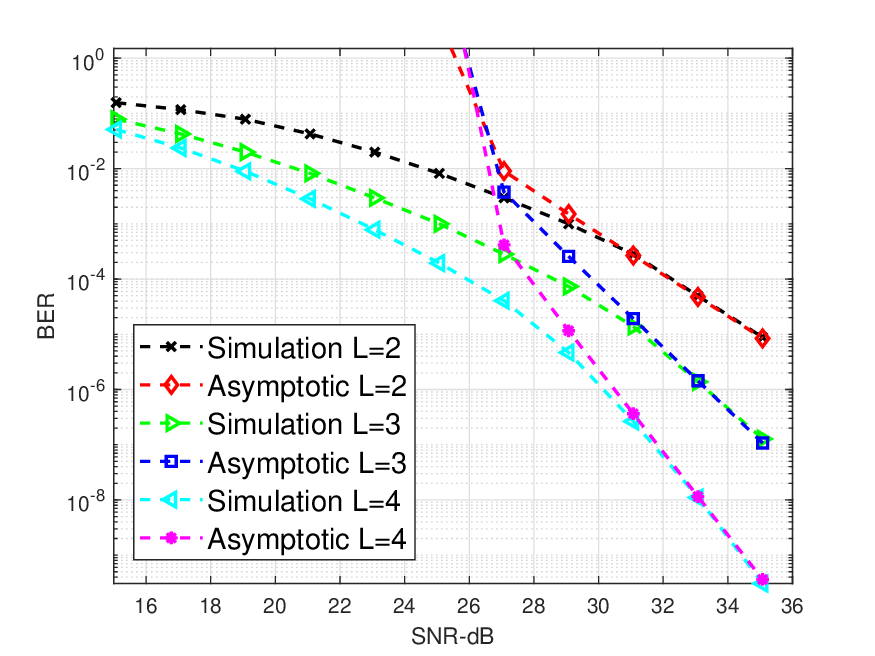}
    \caption{BER versus SNR for i.i.d MG distribution. Asymptotic is in reference to \eqref{SER_MG}.}
    \label{fig:diversity_MG}  
\end{figure}

\begin{figure}[t!]
     \centering
    \includegraphics[width=0.495\textwidth]{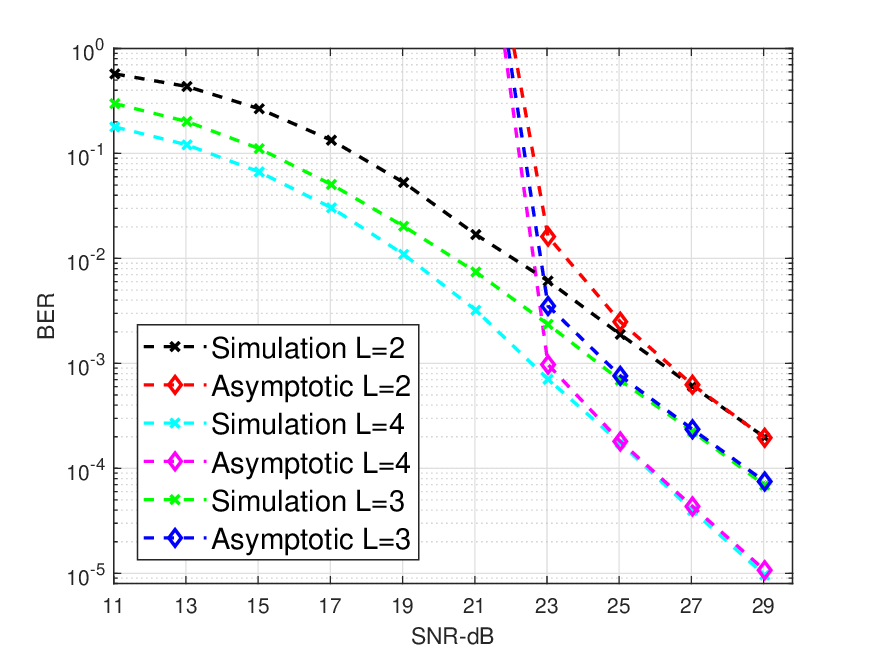}
    \caption{BER versus SNR for i.n.i.d MG distribution. Asymptotic is in reference to \eqref{eq:SER_nonIID_closedform}.}
    \label{fig:diversity_MG_niid}  
\end{figure}

\begin{figure}[t!]
    \centering
    \includegraphics[width=0.495\textwidth]{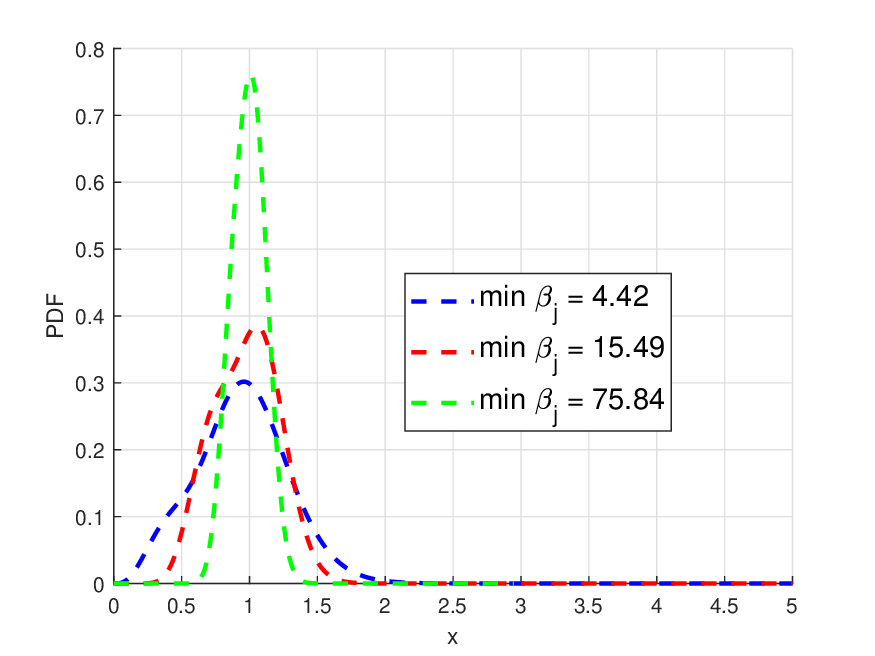}
    \caption{Probability density function of the MG distribution 
    for $N=2$. The curves correspond to different minimum $\beta_j$ values.}
    \label{fig:MG_K2}
\end{figure}

\begin{figure}[t!]
    \centering
    \includegraphics[width=0.495\textwidth]{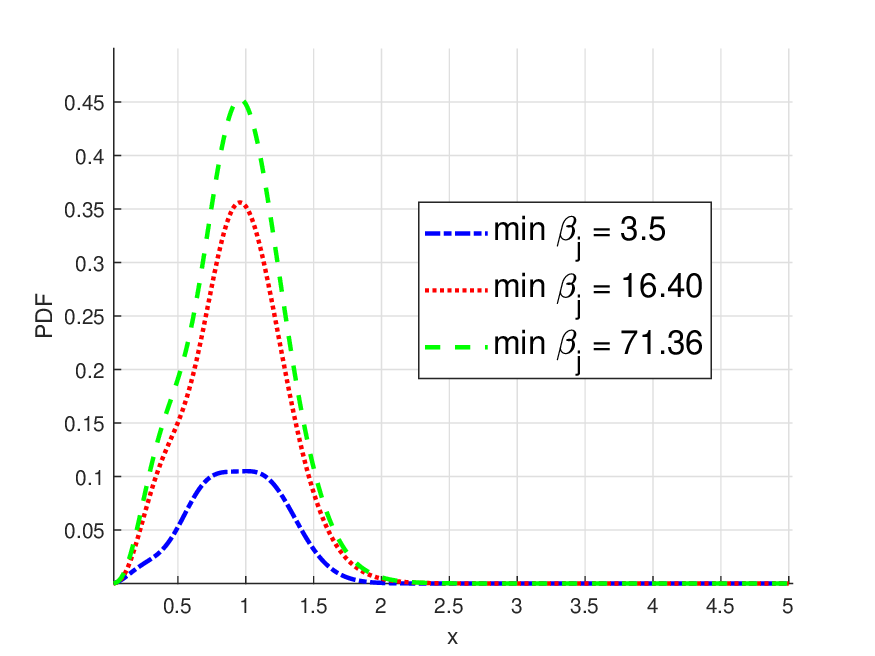}
    \caption{Probability density function of the MG distribution 
    for $N=3$. The curves correspond to different minimum $\beta_j$ values.}
    \label{fig:MG_K3}
\end{figure}

For the THz outdoor scenario, the BER results under i.i.d.\ and i.n.i.d.\ MG fading are presented in Fig.~\ref{fig:diversity_MG} and Fig.~\ref{fig:diversity_MG_niid}, respectively. The MG parameters are taken from Table~\ref{tab:MG}. In the i.i.d.\ case of Fig.~\ref{fig:diversity_MG}, Configuration~1 is employed for all values of $L$. In the i.n.i.d.\ case of Fig.~\ref{fig:diversity_MG_niid}, different configurations are assigned per branch: for $L=2$, Configurations~1 and~2 are used; for $L=3$, Configurations~1, 2, and~3 are adopted; and for $L=4$, Configurations~1 through~4 are applied. The simulations consider $L \!=\! 2, 3, 4$ diversity branches-- for $L=4$, each simulated point is obtained by averaging over $9\times10^{6}$ independent realizations per SNR value-- and demonstrate the adequacy of the asymptotic analysis at high SNR as captured by~\eqref{SER_MG} for the i.i.d case and~\eqref{eq:SER_nonIID_closedform} for i.n.i.d, which represents the asymptotic theoretical curve at high SNR.
In Fig.~\ref{fig:MG_K2} and Fig.~\ref{fig:MG_K3}, the PDF of the MG distribution is plotted for different parameter sets taken from~\cite{papasotiriou2023outdoor}. As observed, increasing the value of $\min(\beta_j)$ leads to a stronger concentration of the PDF around the line-of-sight (LoS) component, indicating improved channel conditions.


\section{Conclusion}
\label{sec:conclusion}
In this paper, we present a diversity analysis for THz communication systems in both indoor and outdoor scenarios. For the indoor case, we recover the diversity results of~\cite{9250676,10003243} using an approximation approach and provide a new exact analysis for the generalized $\alpha$-$\mu$ distribution. For the outdoor scenario, we characterize the diversity behavior using the Laplace transform expression of the PDF of the squared MG distribution. The results from the simulated Monte Carlo runs show excellent agreement with the analytical expressions. Our analysis highlights the impact of the concentration characteristics of the $\alpha$-$\mu$ and MG distributions on diversity gain, demonstrating that tighter clustering of multipath components can enhance signal reliability in dense scattering environments. These insights provide valuable guidelines for the design of efficient and reliable THz communication systems under THz channels conditions incorporating molecular absorption and frequency-dependent path loss.

\appendices
\counterwithin*{equation}{section}
\renewcommand\theequation{\thesection.\arabic{equation}}
\section{Laplace Transform of Squared Mixture of Gamma}
\label{appA}
To compute the Laplace transform of the PDF of the square of MG, we follow the approach of~\cite[Appendix A]{10003243}, where the Laplace transform of the $\alpha$-$\mu$ PDF is derived. The proof steps presented hereafter are similar to those presented in~\cite{10003243}. Using~\eqref{PDF_|h|2_MG}, we write the PDF of $\gamma'=\Upsilon|h|^2$ as
\begin{equation}
    f_{\gamma'}(y)= \sum_{i=1}^N a_i y^{bi-1} \exp{(-c_i\sqrt{y})}
\end{equation}
where $a_i=\alpha_i\Upsilon^{-b_i}/(2\nu^{\beta_i}) , b_i=\beta_i/2 , c_i=\zeta_i/(\sqrt{\Upsilon}\nu)$. The Laplace transform for $\Re\{s\} \geq 0$ is given by 
\begin{align}
    &\mathcal{L}\{f_{\gamma'}(y)\}(s) = \int_0^\infty e^{-sy} f_{\gamma'}(y) \, dy\notag\\
    &= \sum_{i=1}^N a_i \int_0^\infty y^{bi-1} \exp{(-c_i\sqrt{y})}  \exp{(-sy)}\, dy\notag\\
    &= \sum_{i=1}^N a_i \int_0^\infty y^{bi-1}  \exp{(-sy)}G_{0,1}^{1,0} \left[c_i\sqrt{y} \Bigg| \begin{array}{c} \sim \\ 0 \end{array} \right]\, dy,\label{tempa4}
\end{align}
where in order to write equation~\eqref{tempa4}, the exponential function is replaced by its Meijer \(G\)-function representation \(e^{-x} =G_{0,1}^{1,0} \left[x \Bigg| \begin{array}{c} \sim \\ 0 \end{array} \right]\)~\cite{11072420}.
Using~\cite[eq. (07.34.02.0001.01)]{wolfram2025} and after interchanging the order of integration, we rewrite \eqref{tempa4} as
\begin{align}
    \mathcal{L}\left\{f_{\gamma'}\right\}(s) &=  \frac{1}{2\pi j} 
 \sum_{i=1}^N a_i\oint_{\mathfrak{L}_{s_1}} \left(c_i\right)^{-s_{1}} 
 \Gamma\left(s_{1}\right)  \notag\\&\times \int_{0}^{\infty} \exp(-sy)y^{b_i-1-\frac{s_1}{2}}dy \, ds_{1},\label{temp22}
\end{align}
where \( s_{1} \) is a complex variable of integration, and \(\mathfrak{L}_{s_1}\) is a closed contour in the complex plane that encloses all the poles of \(\Gamma(s_{1})\). 
After evaluating the resulting inner integral in \eqref{temp22}, we obtain
\begin{align}
    \mathcal{L}\!\left\{f_{\gamma'}\right\}\!(s) &=\! \frac{1}{2\pi j}\!\sum_{i=1}^N a_i s^{-b_i}  \!\oint_{\mathfrak{L}_{s_1}} \hspace{-2mm}\Gamma\left(s_{1}\right) \Gamma\left(b_i - \frac{s_{1}}{2}\right) \! \left( \frac{c_i}{\sqrt{s}}\right)^{-s_{1}} \hspace{-2.2mm}ds_1 \notag\\
    &= \frac{1}{2\pi j} \sum_{i=1}^N a_i s^{-b_i}  \oint_{\mathfrak{L}_{s_1}^{\dagger}} \Phi(s_1) \, ds_1,\label{tempa6}
\end{align}
where \(\Phi(s_1) = \Gamma\left(s_{1}\right) \Gamma\left(b_i - \frac{s_{1}}{2}\right) \left( \frac{c_i}{\sqrt{s}}\right)^{-s_{1}}
\) and \(\mathfrak{L}_{s_1}^{\dagger}\) is a new contour that appears as the integration over \(y\) deforms \(\Phi(s_1)\). As mentioned in~\cite[Appendix A]{10003243}, the new contour allows to represent \(\mathcal{L}\{f_{\gamma'}\}(s)\) in terms of a meromorphic function analytically defined on the strip \(0 < s_1 < 2 b_i\) and with singularities located at \(s_1 = -t\) and \(s_1 = 2(b_i+t)\), \(\forall t \in \mathbb{N}_0\). The contour \(\mathfrak{L}_{s_1}^{\dagger}\) is chosen to start at \(-\infty + j\xi_1\) and end at \(-\infty + j\xi_2\), for \(-\infty < \xi_1 < \xi_2 < +\infty\), in such a way that \(\mathfrak{L}_{s_1}^{\dagger}\) encloses all the poles of \(\Gamma(s_1)\) in the positive direction. We note that since we are computing $\mathcal{L}\!\left\{f_{\gamma'}\right\}\!(s)$ at $s = \frac{g}{\sin^2\theta}$, $0 \leq \theta \leq \frac{\pi}{2}$ as can be seen in~\eqref{SER_lap} and~\eqref{SER_lap_non_iid}, then $(\frac{c_i}{\sqrt{s}})^{-s_1}$ converges for $\Re(s_1) < 0$ since $\frac{c_i}{\sqrt{s}} = \frac{c_i}{\sqrt{g}} \sin \theta = \frac{\zeta_i}{\nu \sqrt{g\Upsilon}} \sin \theta$ is less than $1$ for large $\Upsilon$, for all $1 \leq i \leq N$.
Now, using Cauchy's residue theorem, equation~\eqref{tempa6} can be expressed as follows~\cite{kreyszig2010advanced}:
\begin{equation}
    \mathcal{L}\{f_{\gamma'}\}(s) = \sum_{i=1}^N a_i s^{-b_i}\sum_{t=0}^{\infty}\text{Res}\left[\Phi(s_1):\{s_1=-t\}\right],
\end{equation}
where \(\text{Res}\left[\Phi(s_1):\{s_1=-t\}\right]\) is the residue of \(\Phi(s_1)\) at the poles \(s_1 = -t\). Evaluating the residues \(\text{Res}\left[\Phi(s_1);\{s_1=-t\}\right]= \lim\limits_{s_1 \to -t}(s+t)\Phi(s_1)\)~\cite[eq. (16.3.3)]{kreyszig2010advanced}, we get:
\begin{equation}
    \mathcal{L}\{f_{\gamma'}\}(s) = \sum_{i=1}^N 
    \frac{\alpha_i \Upsilon^{-\frac{\beta_i}{2}}}{2 \nu^{\beta_i}} \cdot s^{-\frac{\beta_i}{2}} 
    \sum_{t=0}^{\infty} 
    \frac{\Gamma\left(\frac{\beta_i + t}{2}\right) \left(-\frac{\zeta_i}{\sqrt{\Upsilon} \nu \sqrt{s}}\right)^t}{t!}.
\end{equation}
At high values of $\Upsilon$, the argument of the inner series becomes increasingly small, causing the higher-order terms to decay rapidly. As a result, the series is well-approximated by its first term, with the remaining terms contributing negligibly. Thus, we get:
\begin{equation}
    \mathcal{L}\{f_{\gamma'}\}(s) \approx \sum_{i=1}^N 
    \frac{\alpha_i \Upsilon^{-\frac{\beta_i}{2}}\Gamma\left(\frac{\beta_i}{2}\right)}{2 \nu^{\beta_i}} \,s^{-\frac{\beta_i}{2}}. \label{approx_lap}
\end{equation}

\bibliographystyle{IEEEtran}
\bibliography{IEEEabrv,my_bibliography}

\begin{IEEEbiography}[{\includegraphics[width=1in,height=1.25in,clip,keepaspectratio]{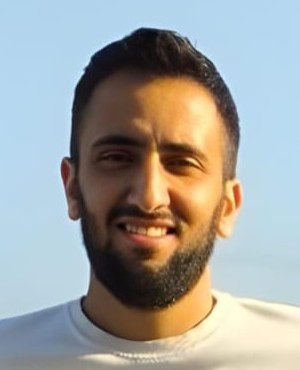}}]{Almutasem Bellah Enad}~(Graduate Student Member, IEEE) received the B.E. degree in Electronics and Communications Engineering from Damascus University in 2023, ranking first in his class for three consecutive academic years and receiving the Award for Academic Excellence. He is currently pursuing the M.E. degree in Electrical and Computer Engineering at the American University of Beirut (AUB), Lebanon. His research interests include signal processing for wireless communications, with a particular emphasis on performance analysis of data detection in high-frequency communication systems. During his master’s studies, he has served as a Teaching Assistant in the Department of Electrical and Computer Engineering at AUB. He serves as a reviewer for IEEE conferences and journals. In the summer of 2025, he was a Student Visiting Researcher with the Information Science Laboratory at King Abdullah University of Science and Technology (KAUST).

\end{IEEEbiography}


\begin{IEEEbiography}[{\includegraphics[width=1in,height=1.25in,clip,keepaspectratio]{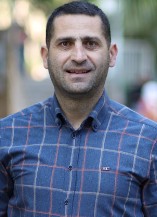}}]{Jihad Fahs}~(Member, IEEE) received the "Diplôme d'Ingénieur en Génie Électrique et Électronique" from the Lebanese University in 2008, and his M.E. and Ph.D. degrees in Electrical and Computer Engineering from the American University of Beirut in 2010 and 2016 respectively. In 2013, he received the CNRS Ph.D. award from the Lebanese National Council for Scientific Research. In 2022, he joined the American University of Beirut where he is currently an assistant professor at the Electrical and Computer Engineering Department. His research interests are in information and estimation theory, wireless communications, and non-standard heavy-tailed probability models. 
\end{IEEEbiography}


\begin{IEEEbiography}[{\includegraphics[width=1in,height=1.25in,clip,keepaspectratio]{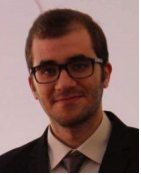}}]
{Hakim Jemaa}~(Graduate Student Member, IEEE) received the Diplôme d’Ingénieur degree from École Polytechnique de Tunisie. He is currently pursuing the Ph.D. degree in electrical and computer engineering with the Communication Theory Lab, within the Computer, Electrical and Mathematical Sciences and Engineering Division, King Abdullah University of Science and Technology (KAUST).
His research interests include signal processing for wireless communications, detection, error correction, and MIMO communications for THz-band systems.
\end{IEEEbiography}


\begin{IEEEbiography}[{\includegraphics[width=1in,height=1.25in,clip,keepaspectratio]{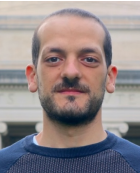}}]
{Hadi Sarieddeen}~(Senior Member, IEEE) received the Ph.D. degree in electrical and computer engineering from the American University of Beirut (AUB) in 2018, where he is currently an Assistant Professor with the Department of Electrical and Computer Engineering.
He held postdoctoral positions with King Abdullah University of Science and Technology (KAUST) and the Massachusetts Institute of Technology (MIT) from 2019 to 2022. In the summer of 2025, he was an Academic Guest with ETH Zurich. His research interests include signal processing for communications, with a focus on THz-band systems, physically consistent signal processing, error correction coding, and semantic communications. He currently serves as an Editor for \emph{IEEE Communications Letters}.
\end{IEEEbiography}


\begin{IEEEbiography}[{\includegraphics[width=1in,height=1.25in,clip,keepaspectratio]{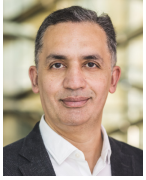}}]
{Tareq Y. Al-Naffouri}~(Fellow, IEEE) received the B.S. degrees in mathematics and electrical engineering (with first honors) from King Fahd University of Petroleum and Minerals, Saudi Arabia, the M.S. degree in electrical engineering from the Georgia Institute of Technology, and the Ph.D. degree in electrical engineering from Stanford University, Stanford, CA, USA, in 2004. 
He was a Visiting Scholar with the California Institute of Technology, Pasadena, CA, USA, in 2005 and during the summer of 2006, and a Fulbright Scholar with the University of Southern California, Los Angeles, CA, USA, in 2008. He is currently a Professor with the Electrical Engineering Department, King Abdullah University of Science and Technology (KAUST).
His research interests include sparse, adaptive, and statistical inference and learning, with applications to wireless communications, localization, smart cities, and smart health. He has authored over 370 journal and conference publications and holds 24 issued or pending patents. 
Dr. Al-Naffouri has received several awards, including the IEEE Education Society Chapter Achievement Award (2008), the Almarie Award for Innovative Research in Communications (2009), the AbdulHameed Shoman Prize for Innovative Research in IoT (2022), and the Research Excellence Award, Innovation in Economies of the Future, RDIA (2025). He has been an IEEE Fellow since January 2025.
\end{IEEEbiography}

\end{document}